\documentclass[aps,prb,twocolumn,superscriptaddress,amsmath,amssymb,floatfix,showpacs,showkeys]{revtex4}

\pdfoutput=1

\usepackage{graphicx}
\usepackage{amsmath}
\usepackage{amssymb}
\usepackage{units}

\begin{document}
    \title{Competing magnetic interactions in exchange bias modulated films}

    \author{Christine Hamann}
    \affiliation{Institute for Metallic Materials, IFW Dresden, Postfach 270116, D-01171 Dresden, Germany}

    \author{Boris P. Toperverg}
    \affiliation{Department of Physics and Astronomy, Institute for Solid State Physics, Ruhr-University Bochum, D-44780 Bochum, Germany}
    \altaffiliation{Petersburg Nuclear Physics Institute, Gatchina, St.Petersburg 188300, Russia}
    \altaffiliation{Institute Laue-Langevin, Grenoble, France}

    \author{Katharina Theis-Br\"ohl}
    \affiliation{University of Applied Sciences Bremerhaven, An der Karlstadt 8, D-27568 Bremerhaven, Germany}
   	\altaffiliation{Department of Physics and Astronomy, Institute for Solid State Physics, Ruhr-University Bochum, D-44780 Bochum, Germany}

    \author{Maximilian Wolff}
		\affiliation{Department of Physics, Uppsala University, Box 530, 751 21 Uppsala, Sweden}
    
    \author{Rainer Kaltofen}
    \affiliation{Institute for Integrative Nanosciences, IFW Dresden, Postfach 270116, D-01171 Dresden, Germany}

    \author{Ingolf M\"onch}
    \affiliation{Institute for Integrative Nanosciences, IFW Dresden, Postfach 270116, D-01171 Dresden, Germany}

    \author{Jeffrey McCord}
    \affiliation{Institute for Metallic Materials, IFW Dresden, Postfach 270116, D-01171 Dresden, Germany}

    \author{Ludwig Schultz}
    \affiliation{Institute for Metallic Materials, IFW Dresden, Postfach 270116, D-01171 Dresden, Germany}
    
\date{\today}
\begin{abstract}

The magnetization reversal in stripe-like exchange bias patterned $\rm Ni_{81}Fe_{19}/IrMn$ thin films was investigated by complementary inductive and high resolution magneto optical magnetometry, magneto optical Kerr microscopy, and polarized neutron reflectometry to clarify the effects of competing interfacial exchange bias and lateral interface contributions. Structures of varying ferromagnetic layer thickness and stripe period were analyzed systematically at the frozen-in domain state of oppositely aligned stripe magnetization. For all samples the mean magnetization of the magnetic hybrid structures was found to be aligned nearly orthogonally with respect to the stripe axis and the set exchange bias direction. Due to the interaction of interfacial coupling, exchange, and magneto-static energy contributions, the opening angle of neighboring stripe magnetizations increases with decreasing ferromagnetic layer thickness and increasing stripe period. The experimental observations are in agreement with an earlier proposed model for designing micro-patterned exchange bias films. 

\end{abstract}

\maketitle

\section{introduction}
Extended hybrid thin films with locally modulated static and dynamic magnetic properties have recently gained considerable attention from a fundamental as well as application related point of view (see Refs.~\onlinecite{fassbender08,mccord08d} and references therein). In such films lateral magnetic patterns are imprinted without a significant modification of the surface structure  \cite{chappert98}. A successful approach  is the local modification of magnitude and/or direction of the exchange bias (EB) effect in ferromagnetic/antiferromagnet (F/AF) bilayers. In a recent study \cite{theis-brohl06} we locally altered the
sign of the EB effect in a CoFe/IrMn film by Ion Beam Induced Magnetic Patterning (IBMP) \cite{ehresmann04,fassbender08}. With
this method we produced a periodic array of stripes of equal width and alternating unidirectional anisotropy. Hence, at zero external fields a domain pattern with antiparallel alignment of the magnetization in neighboring stripes was expected so that the net unidirectional anisotropy is totally compensated. However, it was found that the stripe magnetization is not aligned completely antiparallel to each other, but significantly tilted with respect to the stripe axis. This results in an co-existence of different uniaxial anisotropies with the effective easy axis of magnetization aligned along or perpendicular to the stripe axis for the case of parallel, respectively antiparallel alignment of magnetization. The reason for the combination of two types of anisotropy is caused by the competition between the lateral ferromagnetic exchange interaction within the F layer and the alternating interfacial exchange bias field. In a simple theoretical model\cite{theis-brohl06} we have explained
the phenomenon and estimated conditions allowing for the existence of a purely antiparallel orientation of magnetization. 
Whereas the EB exchange field contribution is proportional to the area of AF/F interface, the bulk exchange energy of the F-layer decomposed into a set of antiparallel stripe domains is proportional to the area of domain walls running through the ferromagnetic film thickness. Therefore, the ratio between bulk and interfacial contributions to the total free energy is roughly proportional to the ratio $t_{\rm F}/D_{\rm st}$ of the film thickness $t_{\rm F}$ to the stripe width $D_{\rm st}$. This means, the interfacial EB effect may cause antiparallel configuration of the stripe magnetization only for sufficiently thin F layers. Contrary, thick F layers tend to be homogeneously magnetized. In the range of intermediate values of $(t_{\rm F}/D_{\rm st})\sim |\rm J_{EB}|/J_{\rm F}$, where $\rm J_{\rm F}$ is the ferromagnetic and $\rm J_{EB}$ is the mean value of the interfacial exchange integrals, the system is unstable with respect to a tilt of the stripe magnetization and the coexistence of lateral ferromagnetic and antiferromagnetic order. The resulting tilt angle between the stripe magnetization
directions therefore is assumed to increase with either increasing stripe width or decreasing ferromagnetic film thickness, while keeping the interfacial and bulk ferromagnetic exchange coupling unchanged. However, a symmetric configuration\cite{theis-brohl06} with equally wide stripes and perfectly alternating EB fields only allows for an alignment of net magnetization of $90^\circ$ relative to the stripe axis. This is not the case for asymmetric structures with either unequal stripe widths or different interfacial coupling strengths for neighboring stripes. A variation of those parameters may provide an opportunity to design systems with a designated angle between net uniaxial and unidirectional anisotropy and hence the direction of total magnetization.
The aim of the present study is to verify theoretical expectations for a F/AF bilayer systems with different F layer thickness,
stripe widths and asymmetric interfacial coupling contribution. Therefore, in the present set of experiments we used planar film structures consisting of a continuous soft ferromagnetic layer, which is partially exchange biased by a laterally periodic
antiferromagnetic film stripe array. We studied films with varying ferromagnetic film thickness and stripe width. This enabled
us to investigate the overall remagnetization process in dependence on different EB and F exchange contributions. 
\section{Experimental}

\begin{figure}[htbp]
\includegraphics[width=6.5cm]{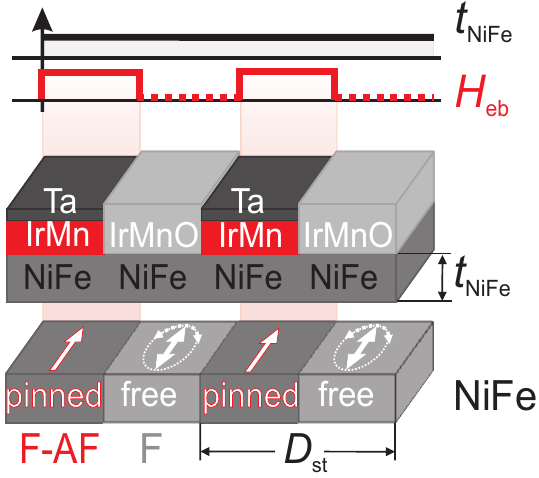}
\caption{\label{fig1} (Color online) Schematics of the thin film geometry with $D_{\rm st}$ being the stripe period. The magnetic and structural properties modulated over the stripe structure are shown.}
\end{figure}

The F/AF samples were prepared as polycrystalline continuous $\rm Ta(4~nm)/Ni_{81}Fe_{19}(t_{\rm F})/Ir_{23}Mn_{77}(t_{\rm AF})/Ta(3~nm)$ films deposited on oxidized Si-wafers by dc-magnetron sputtering. In contrast to earlier studies\cite{theis-brohl06} with CoFe as F layer, NiFe was used due to the ideal soft magnetic properties of low induced uniaxial anisotropy and absence of magnetic anisotropy dispersion, which is common to $\rm CoFe$ films. In order to obtain a 2-dimensional lateral modulation of the unidirectional exchange anisotropy $K_{\rm ud}$, a superimposed stripe array of IrMn and $\rm IrMnO_{\rm x}$ was generated. This was obtained by subjecting the layer stacks to a subsequent lithography patterning and oxidation process (for more details on the preparation procedure see Ref.~\onlinecite{hamann08}). Hereby, $\rm IrMn$ is antiferromagnetic and able to introduce an exchange bias field to the F layer, while $\rm IrMnO$ is non-magnetic. At a temperature of $250~^\circ$C, which is above the blocking temperature of the system, an external field $H_\mathrm{ann}$ of $\unit[1]{T}$ was applied to set the unidirectional anisotropy direction of the F-AF stripes to be aligned along the stripe axis. The resulting exchange bias patterned film of nominal $\rm Ta(4~nm)/Ni_{81}Fe_{19}(t_{\rm F})/Ir_{23}Mn_{77}-IrMnO_{x}(t_{\rm AF})/Ta-IrMnO_{x}(t_{\rm AFO_{x}})$ comprises exchange biased (F-AF) stripes and unbiased ferromagnetic stripes (F) through an underlying extended ferromagnetic layer (Fig.~\ref{fig1}). The stripe period $D_{\rm st}$ was chosen to be $4~\mu$m and $12~\mu$m to vary the amount of inter-stripe interfaces and being still within length scales suitable for neutron scattering experiments.
The relative contribution of ferromagnetic exchange energy was varied by adjusting the ferromagnetic layer thickness $t_{\rm F}$
to be either $20$~nm or $30$~nm, while the antiferromagnetic layer thickness $t_{\rm AF}\geq7$~nm was always kept well above the critical thickness for the occurence of exchange bias~\cite{mccord04}. 

To clarify the magnetization reversal process, we combined different complementary magnetization characterization methods. First, we employed inductive magnetometry to map the macroscopic magnetic properties. Secondly, by means of high resolution Kerr microscopy, the microscopic reversal mechanisms were studied quasi-statically. Thirdly, neutron scattering measurements have been conducted to extract quantitative information on the magnetization distribution. Polarized neutron reflectometry (PNR) has been proven to be a powerful tool to quantitatively map the vectorial magnetization of periodically patterned media~\cite{theis-brohl06,temst05,zabel07}. Last, high resolution MOKE magnetometry in longitudinal and transverse sensitivity was used to obtain quantitative information on the local magnetization within each magnetic phase orientation with submicron resolution. 

\begin{figure}[htbp]
\includegraphics[width=8cm]{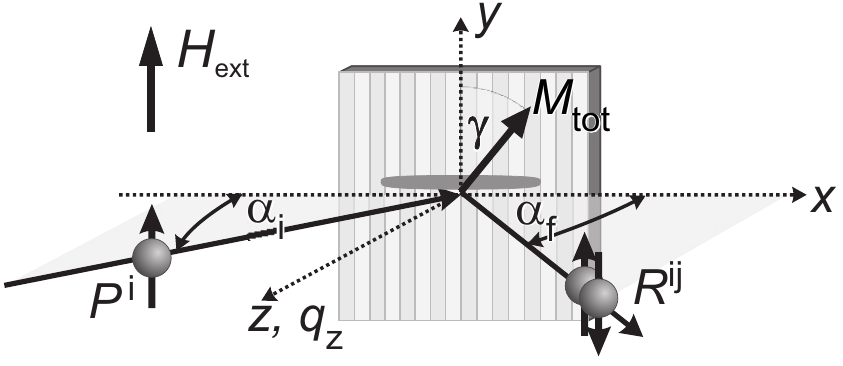}
\caption{\label{fig2} Scattering geometry for polarized neutron experiments.}
\end{figure}

The PNR experiments were carried out at the ADAM reflectometer~\cite{schreyer98,wolff07} at the ILL Grenoble using a fixed cold neutron wavelength of $\unit[0.441]{nm}$. To apply external fields $H_{\rm ext}$ along the stripe axis, an electromagnet with the field direction parallel to the sample surface and parallel to the neutron polarization as well as perpendicular to the scattering plane was used (see Fig.~\ref{fig2}). The incoming neutron polarization $P_i$ was either parallel ($+$) or antiparallel ($-$) to the external magnetic field guiding the neutron polarization. Four specular reflectivities: two non-spin-flip (NSF), $\mathcal{R}^{++}$ and $\mathcal{R}^{--}$, and two spin-flip (SF), $\mathcal{R}^{+-}$ and $\mathcal{R}^{-+}$, were measured.
Additionally, the specular reflection and off-specular scattering, including Bragg diffraction from lateral patterns were measured with a position sensitive detector (PSD) over a broad range of incident and scattering angles, respectively $\alpha_i$ and $\alpha_f$. In the latter measurements the scattered intensities were recorded for altered incident polarization, while the final spin state of the neutrons was not discriminated.
NSF reflectivities are sensitive to the magnetization vector projections parallel to the polarization vector\cite{zabel07} (y-axis in Fig.~\ref{fig2}), while magnetization components normal to this direction cause SF reflection. Due to the occurrence of magnetic domains, either random or periodic, the magnetization direction may vary over the film surface and the measured SF and NSF reflected intensities are a result of averaging over all magnetization directions. In general, such an averaging is not a trivial procedure, because the reflection amplitudes are quite complicated non-linear functions of the optical and 
magnetic potential. Nevertheless, in most cases of practical interest, the averaging can be substantially simplified, taking into account that the specular reflection is caused by the mean optical potential averaged over the coherence ellipsoid. Its projection onto the sample surface is indicated in Fig.~\ref{fig2}. At grazing incidence the longest axis of the ellipsoid is parallel to the line of intersection between reflection plane and the sample surface. This axis may extend up to the sub-millimeter scale, while the two other principle axes, one within and the other normal to the surface, amount only to a few tens of nanometers.
The coherence ellipsoid consists of a narrow stripe, comprising only a very small fraction of the sample, the total volume of which is covered with a number of ellipsoids. Hence, the observed reflection signal is an incoherent sum of intensities coherently reflected from different parts of the sample. In each of these parts the magnetization, which is averaged over a particular coherence ellipsoid, is tilted by an angle $\gamma$ varying across the surface. 
On the other hand, the longest axis of the coherence ellipsoid crosses a number of domains or structural elements in micro-patterned films. Therefore the optical potential expressed via the scattering length density (SLD) $Nb$ is almost independent of the position of the coherence ellipsoid. This allows to approximate the NSF and SF reflectivities with the following equations\cite{toperverg99,toperverg02,zabel07}:
\begin{eqnarray}
	\label{reflectpp}
	{\mathcal R}^{\pm\pm}&=&\frac{1}{4}\{|R_+|^2\langle(1\pm\cos\gamma)^2\rangle+
	|R_-|^2\langle(1\mp\cos\gamma)^2\rangle\nonumber \\
	&+&2\Re{(R_+R_-^*)}\langle\sin^2\gamma\rangle\},\\
	\label{reflectpm}
	{\mathcal R}^{\pm\mp}&=&\frac{1}{4}|R_+-R_-|^2\langle\sin^2\gamma\rangle.
\end{eqnarray}
Here\footnote{These equations are valid for ideal polarization and polarization analysis. If the sample is in saturation and the magnetization is parallel to the polarization vector then $\mathcal{R}^{++}=|R_+|^2$, $\mathcal{R}^{-\,-}=|R_-|^2$ and
$\mathcal{R}^{+-}=\mathcal{R}^{-+}=0$.} $R_\pm(q_z)$ are reflection amplitudes for positive and, correspondingly, negative neutron spin projections onto the mean magnetic induction within a coherence ellipsoid.
The functions $\langle\cos\gamma\rangle$ and $\langle\sin^2\gamma\rangle$ are averaged over different coherence ellipsoids covering the sample surface. The reflection amplitudes $R_\pm(q_z)$ depend upon the transverse wave vector transfer $q_z$ and can readily be calculated with the conventional Parratt\cite{parratt54} routine. Due to the birefringence effect~\cite{toperverg99} in magnetic films each of the amplitudes is determined by its own SLD: $Nb_\pm=\overline{(Nb)}_n\pm\overline{(Nb)}_m\langle\cos(\Delta\gamma)\rangle_\mathrm{coh}$, where $\overline{(Nb)}_{n,m}$ are the mean values of the nuclear (n) and magnetic (m) SLD averaged over the elements of lateral structure within the coherence range. If within this range the magnetic moments of the domains experience deviations by an angle $\Delta\gamma$ from
the direction of the mean magnetization, the magnetic SLD is additionally reduced by the factor $\langle\cos(\Delta\gamma)\rangle_\mathrm{coh}$.
This factor, along with two other parameters for incoherent averaging $\langle\cos\gamma\rangle^\mathrm{inc}$
and $\langle\sin^2\gamma\rangle^\mathrm{inc}$ is determined by fitting the experimental reflectivities measured at different points of the magnetic reversal loop of the magnetic hybrid structure to the theoretical model.

\section{Results}
\subsection{Magnetometry and Kerr microscopy}

\begin{figure}[htbp]
\includegraphics[width=6.5cm]{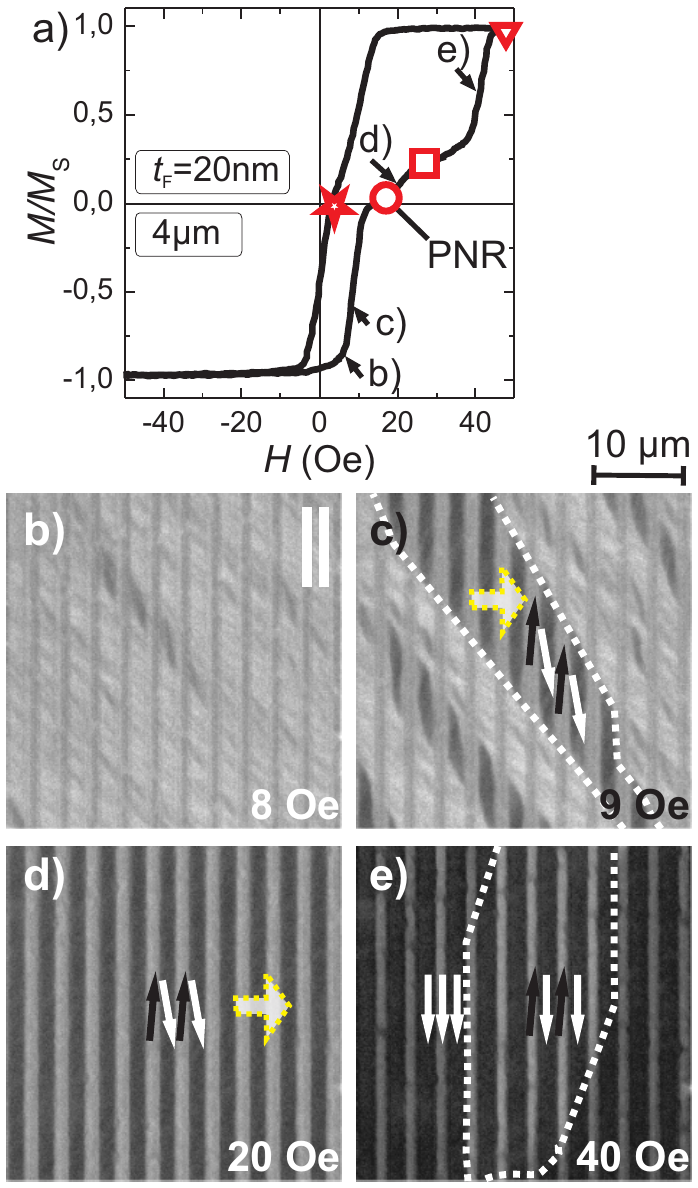}
\caption{\label{fig3} (Color online) Magnetic hysteresis loop (a) of $\rm NiFe(20~nm)/IrMn-IrMnO/Ta-IrMnO$ exchange bias patterned film with $D_{\rm st}=~4~\mu$m. The symbols represent the magnetic field for the PNR measurements (Fig.~\ref{fig4}). Corresponding longitudinal Kerr images along the magnetization loop are depicted in (b)-(e). The external magnetic field values are indicated.}
\end{figure}

In Fig.~\ref{fig3}a) an inductively measured hysteresis loop obtained with the field parallel to the stripe axis
for the film with a thickness of $t_{\rm F}=20$~nm and with a stripe period of $D_{\rm st}=4~\mu$m is shown. For clarification, selected corresponding high-resolution Kerr microscopy magnetic domain images are depicted in Figs.~\ref{fig3}(b-e).
The magnetization loop exhibits a two-step switching process, which is typical for such a magnetic hybrid two-phase system. Starting from negative magnetic fields, first, below $10~$Oe the free F stripe magnetization starts to ripple [Fig.~\ref{fig3}b)], followed by the reversal of magnetization in the stripes by head-on-domain wall motion [Fig.~\ref{fig3}c)]. At the hysteresis step, a rather continuous change of magnetization is observed, indicating magnetization rotation and
relaxation along the positive field direction. Yet, the magneto optical contrast of the domain image [Fig.~\ref{fig3}d)] implies
antiparallel alignment of the two stripe magnetizations. During the second step at $40~$Oe, the inter-stripe domain walls move into the neighboring exchange biased stripes. The pinned magnetization is finally reversed by head-on-domain wall motion [Fig.\ref{fig3}e)]. During the remagnetization process, stripes of the same magnetic properties reverse in a correlated manner, forming so called quasi-domains\cite{mccord05,hamann08} or hyper-domains\cite{theis-brohl08}.

However, the true vectorial magnetization orientation on the hysteresis step, i.e. the plateau region with oppositely stripe
magnetization, is not easily determined from the longitudinal magnetization and domain images. Moreover, due to the optical
resolution limit of about 300~nm, possible intra-stripe variations of magnetization with frequencies below the optical
resolution would not be detected. 

\subsection{Field dependent specular neutron reflectivity}

\begin{figure}[htbp]
\includegraphics[width=8cm]{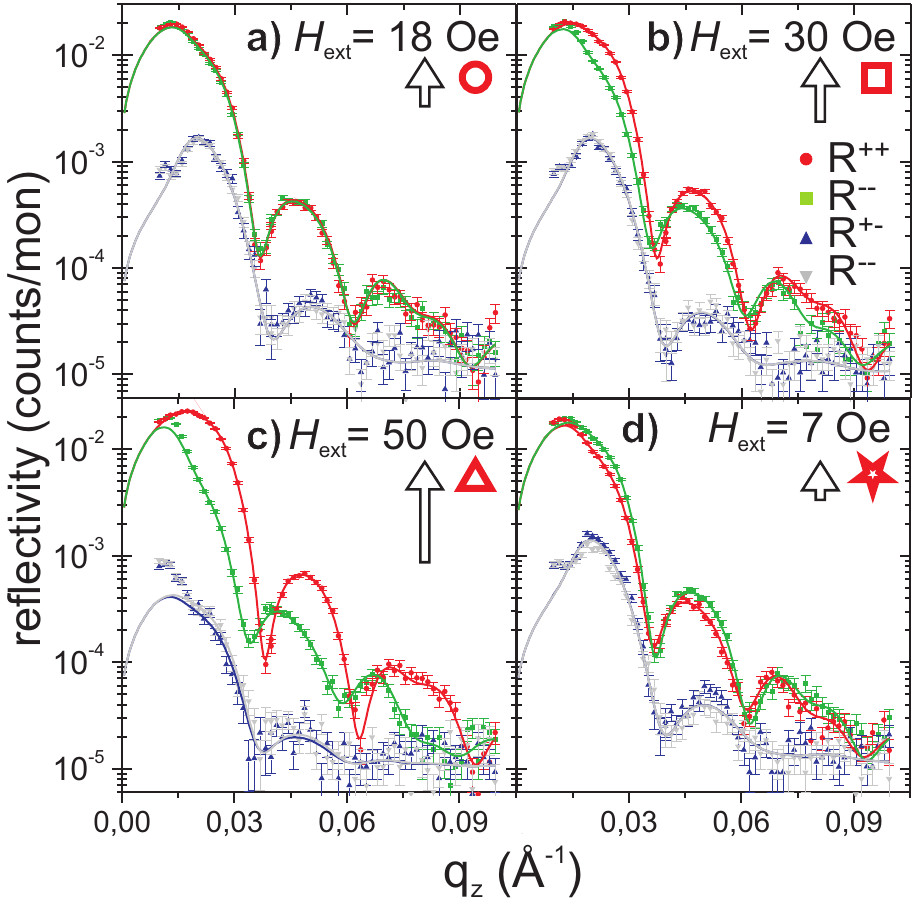}
\caption{\label{fig4} (Color online) PNR specular reflectivity data (symbols) of $\rm NiFe(20~nm)/IrMn-IrMnO/Ta-IrMnO$ with $D_{\rm st}=~4~\mu$m and according fits (lines). The external magnetic field is aligned along the stripe axis and opposite to the unidirectional anisotropy. The magnetic field values and positions along the magnetization loop marked with the circle,
square, triangle, and star correspond to those in Fig.~\ref{fig3}.}
\end{figure}

In order to obtain quantitative information on the magnetic states of the film, specular PNR measurements were performed at selected external magnetic fields. Prior to the measurements, the sample was saturated in negative field direction. Thereafter the magnetic field was increased along the ascending loop branch opposite to the direction of imprinted unidirectional anisotropy. For measurements along the recoil branch, the magnetic field was lowered from positive saturation to zero field. In
Fig.~\ref{fig4} the experimental specular reflectivities are plotted together with the results of the fit to the theoretical
model (see below). The applied field values $H_{\rm ext}$, at which the PNR curves were obtained, are marked in the magnetization loop in Fig.~\ref{fig3}a). The relevant magnetic states, namely the magnetization configurations at the hysteresis step, in positive saturation, and close to the coercive field along the recoil branch, are probed.

At an external field of $H_{\rm ext}=~18~$Oe aligned opposite to the imprinted unidirectional anisotropy, the dark and bright contrast in the magnetic domain images displayed in Fig.~\ref{fig3}d) indicate a mostly antiparallel alignment of magnetization with the free uncoupled stripe magnetization aligned along the magnetic field. The corresponding NSF intensities ($\mathcal{R}^{++}, \mathcal{R}^{-\,-}$) in Fig.~\ref{fig4}a) are almost indistinguishable. This indicates that either the net magnetization is equal to zero or is tilted perpendicularly with respect to the external magnetic field guiding the neutron polarization. Zero total magnetization averaged over the neutron coherence range would be achieved for the state, where the magnetization vectors of neighboring stripes are equal in magnitude, but reversely directed. However, for the magnetization configuration shown in Fig.~\ref{fig4}a), the considerable SF intensities reveal a magnetization component perpendicular to the stripe and the axis of neutron polarization. This component in principle could be related to homogeneous
magnetization rotation across the whole sample as sketched in Fig.~\ref{fig3}d) or within (quasi-)domains larger than the coherence length as indicated in Fig.~\ref{fig3}e). Thus, in agreement with the investigation on fully exchange coupled stripe structures \cite{mccord05,theis-brohl06} an immediate qualitative analysis of Fig.~\ref{fig4}a) and Fig.~\ref{fig3}d) already shows that the magnetization of neighboring stripes at the intermediate hysteresis step is not aligned oppositely, but tilted with respect to the orientation of field and exchange bias.

Further increasing the field ($H_{\rm ext}=30$~Oe) causes a splitting between the NSF reflectivities as shown in Fig.~\ref{fig4}b). This indicates an increased magnetization projection onto the external field direction. At $H_{\rm ext}=50$~Oe the sample is saturated and the NSF reflectivities exhibit the maximum NSF splitting. The intensity of the SF channels is reduced to a minimum value remaining due to the non-perfect neutron polarization limited by the polarizing and analyzing efficiencies. Away from saturation, approaching the coercive field value along the recoil branch (Fig.~\ref{fig4}d), at $H_{\rm ext}=~7~$Oe, the NSF splitting is strongly reduced accompanied by an increase of the SF intensities. Consequently, the situation with nearly compensated longitudinal magnetization components and a considerable transverse magnetization component is comparable to the configuration along the forward branch at the hysteresis step [$H_{\rm ext}=18~$Oe, Fig.~\ref{fig4}a)]. The slightly higher intensity of $\mathcal{R}^{--}$ proves that the mean magnetization is not compensated along the stripe axis as depicted in Fig.~\ref{fig3}a (marked by $\star$). The reason for the noticeable deviation of the magnetometric and PNR results is related to the steep slope in the magnetization loop at this field value. By that, already a small deviation in the field adjustment of $-0.5~$Oe results in a magnetic configuration, where the mean magnetization is negative and thereby would thus comply with the PNR observation. The results of the qualitative analysis of the PNR data are well consistent with the observations by high resolution Kerr microscopy, providing a general idea about the scenario of remagnetization in the system. 

\begin{figure}[htbp]
\includegraphics[width=6 cm]{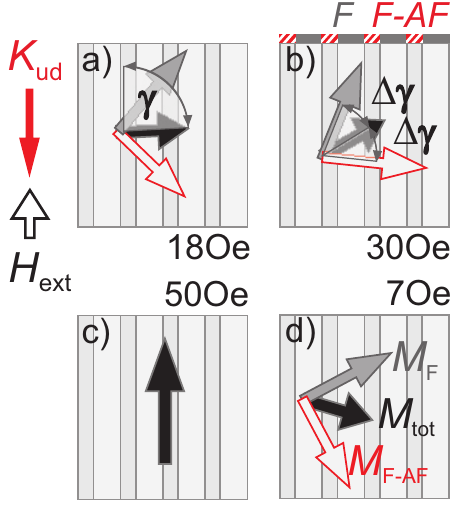}
\caption{\label{fig5} (Color online) Sketch of the alignment of mean total magnetization $M_{\rm tot}$ of $\rm NiFe(20~nm)/IrMn-IrMnO/Ta-IrMnO$ with $D_{\rm st}=~4~\mu$m derived from PNR fits (cf. \ref{fig4}). The different stripe properties are marked by different colors with the exchange biased stripes in hatched red. The free NiFe stripes are labeled by dark gray bars. The applied field and direction of the unidirectional anisotropy $K_{\rm ud}$ of the exchange biased stripes are indicated.}
\end{figure}

Quantitative information on the magnetic states during the remagnetization process are extracted from PNR data by fitting the data to the theoretical model using Eq.~\ref{reflectpp} and Eq.~\ref{reflectpm}. Herewith, all four reflectivities are fitted simultaneously in one cycle, providing excellent agreement\footnote{A few points close to the origin deviate from the model curve due to contamination by the direct beam at very low angles of incidence. Those few points were excluded from the fit.} between data and model calculations. The initial parameters, including polarization efficiencies, layer thicknesses obtained from Transmission Electron Microscopy (TEM) investigations (not shown here), nuclear and magnetic SLDs were refined by fitting the saturated magnetization state fixing the magnetic parameters: $\langle\cos\gamma\rangle=1$, $\langle\sin^2\gamma\rangle=0$, and $\langle\cos(\Delta\gamma)\rangle_\mathrm{coh}=1$. 
Via the PNR fitting the thickness values of the lateral alternating layer stack were corrected to 
$\rm Si/SiO/Ta(2~nm)/Ni_{81}Fe_{19}(20~nm)/Ir_{23}Mn_{77}-IrMnO_{x}(7~nm)/Ta-IrMnO_{x}(5~nm)/IrMnO_{x}(7~nm)$. The deviations in individual layer thicknesses relative to the nominal values are attributed to the patterning and annealing process.
In order to account for the interdiffusion of NiFe during the oxidation procedure, two intermixing layers between NiFe and Ta, respectively IrMn, were introduced into the model. Thereby, the total magnetic layer thickness remains constant ($t_{\rm
F}=20~$nm), but due to the intermixing, the nuclear and magnetic SLDs of the interdiffusion regions become an average of the involved layers. The final layer stack in the model remains with a magnetically unchanged NiFe layer of $t_{\rm F,NiFe}=11~$nm as a core, which is sandwiched between two layers of reduced magnetic moment and with the thickness of $t_{\rm F,NiFe-Ta}=6$~nm and $t_{\rm F,NiFe-IrMn}=3~$nm for the alloyed NiFe-Ta, respectively the NiFe-IrMn layer.
The deduced orientation of the mean magnetization $M_{\rm tot}$ relative to the neutron polarization and stripe axis and its absolute value is found by fitting of the three parameters $\langle\cos\gamma\rangle$, $\langle\sin^2\gamma\rangle$, and $\langle\cos(\Delta\gamma)\rangle_\mathrm{coh}$ (see Fig.~\ref{fig2} and Fig.~\ref{fig5}), fixing all other parameters to the values obtained in magnetic saturation. The reduction of the mean optical potential due to magnetization fluctuations around the net magnetization direction, e.g. domains with dimensions below the neutron coherence length, is expressed by a dispersion angle $\Delta\gamma$ relative to its mean value $\gamma$. Assuming homogeneous intra-stripe magnetization (discussed in Sec.~\ref{sec:offSpec}) the angle $\Delta\gamma$ may vary periodically from one stripe's domain to the next. Therefore, the magnetization $M_{\rm F}$ of the free uncoupled layer fraction is assumed to be tilted at an angle $\gamma+\Delta\gamma$, while the exchange biased stripe magnetization $M_{\rm F-AF}$ has the angle $\gamma-\Delta\gamma$, or vice versa, with respect to the polarization direction. In principle, the deviation angle $\Delta\gamma$ can also be randomly distributed within each stripe as well as across neighboring stripes due to small magnetic ripple domains. 
While specular PNR cannot distinguish between periodic and random deviations of the magnetization directions and only determines a reduction factor $\langle\cos(\Delta\gamma)\rangle_\mathrm{coh}\leq1$ of the mean magnetic optical potential, the missing details are obtained by analyzing the off-specular scattering as discussed below. Moreover, specular PNR yields not only the reduction of the mean magnetization, but also its mean direction, or more precisely, the value $\langle\cos\gamma\rangle$ averaged over the whole sample surface. Such averaging includes, in particular, quasi-domains as seen in Fig.~\ref{fig3}c) and Fig.~\ref{fig3}e). Moreover, from the parameter $\langle\sin^2\gamma\rangle=1-\langle\cos^2\gamma\rangle$ a magnetic dispersion $\langle\cos^2\gamma\rangle-\langle\cos\gamma\rangle^2\geq0$ in the longitudinal projections of the magnetization averaged over the coherence length can be estimated. In the case of pure coherent rotation of the mean magnetization the dispersion equals zero, while alternatively, the condition $\langle\cos^2\gamma\rangle\geq\langle\cos\gamma\rangle^2$ is fulfilled.

Figure~\ref{fig5} displays the evolution of $M_{\rm tot}$ during the magnetization reversal process. For the magnetization configuration at the plateau close to the coercive field at $H_{\rm ext}=18$~Oe the stripe structure the vector of net magnetization is tilted at the angle $\gamma=87^\circ$, i.e. almost orthogonal to the stripe and neutron polarization axis. Furthermore, from the reduction factor $\langle\cos\Delta\gamma\rangle$ of the mean magnetization, i.e. the according dispersion angle $\Delta\gamma\approx50^\circ$, one can deduce that the magnetization vector of the F stripes $M_{\rm F}$ is
reversed and tilted with respect to the direction of the external field by an angle of $37^\circ$. The magnetization in the
exchange biased stripes $M_{\rm F-AF}$ still points opposite to the applied field direction enclosing an angle with the field of
about $137^\circ$. An asymmetric magnetic configuration evolves. By increasing the magnetic field along the intermediate plateau of the magnetization loop, at $H_{\rm ext}=18$~Oe, the average magnetization is further rotated along the field direction now enclosing an angle of $\gamma=59^\circ$. The according magnetization fluctuation of $\Delta\gamma=37^\circ$ is reduced as the free stripe magnetization is now aligned almost along the field and the exchange biased fraction is further rotated toward the magnetic field. At $H_{\rm ext}=50~$Oe the film is saturated. 
Decreasing the field again, the exchanged biased stripes relax toward the unidirectional anisotropy direction, which leads to the mean magnetization aligned along $\gamma=107^\circ$. Hereby, the rotation occurs along the same semicircle, to the right, as for the forward field direction. This is a feature typical for exchange biased bi-layers, especially for situations with misalignment of the unidirectional anisotropy with respect to the external field\cite{beckmann03,mccord08a}. In the free stripes
the vector $M_{\rm F}$ is still pointing along the magnetic field direction, spanning an angle of $\Delta\gamma=45^\circ$ to $M_{\rm tot}$. Nevertheless, due to the coupling via the stripe interface the transverse magnetization component, $M_{\rm F-AF}$ mediates the $M_{\rm F}$ rotation direction.
\subsection{Off-specular scattering} \label{sec:offSpec}

\begin{figure}[htbp]
\includegraphics[width=8cm]{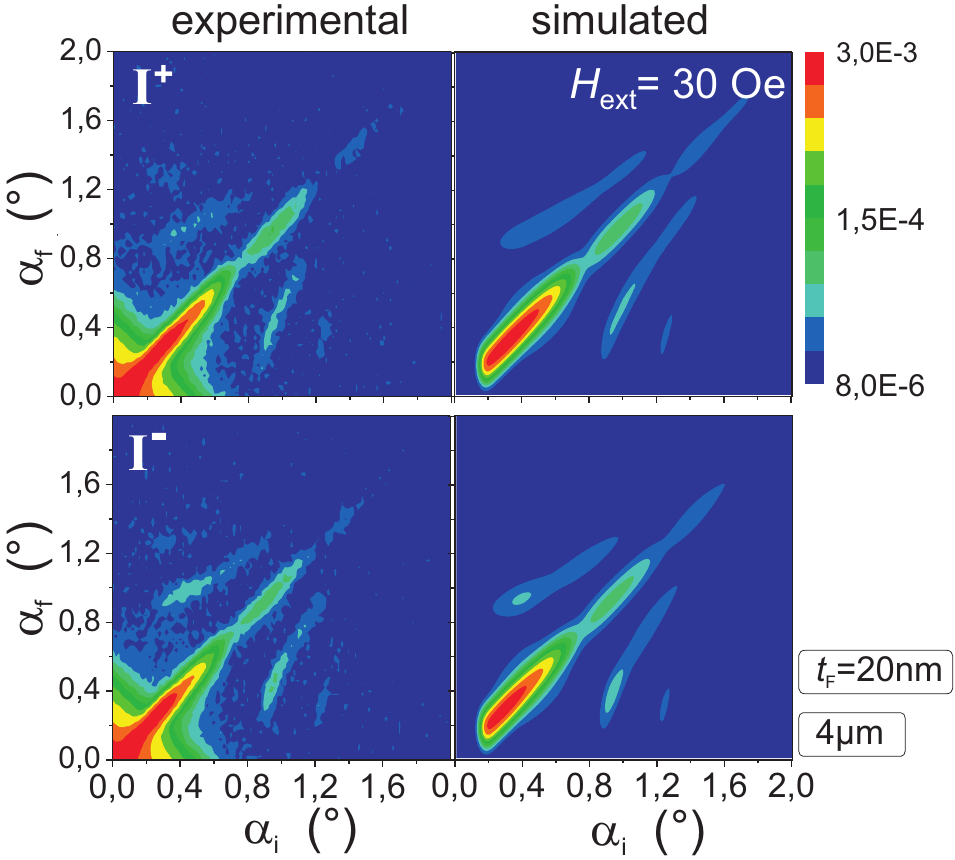}
\caption{\label{fig6} (Color online) Experimental (left column) and
simulated (right column) polarized neutron scattering intensity maps
of $\rm NiFe(20~nm)/IrMn-IrMnO/Ta-IrMnO$, $D_{\rm st}=~4~\mu$m. The
intensities of $R^+$ and $R^-$ are plotted as function of the
incident and scattering angle $\alpha_{\rm i}$ and $\alpha_{\rm f}$,
respectively. The intensity is plotted in logarithmic scale.}
\end{figure}

In order to confirm the model, which was used to fit the specular reflectivities and also to test the hypothesis of a homogeneous intra-stripe magnetization component at the step of the magnetization loop, neutron specular reflection and off-specular scattering was measured with the position sensitive detector. In this case the neutron beam impinging on the sample surface was polarized along with or opposite to the mean magnetization direction. No analysis of spin states of scattered neutrons was applied. The measured intensities for each direction of the incident polarization were recollected into the maps displayed in the left column of Fig.~\ref{fig6}, indicating the intensity $I^+$ and $I^-$ distributions versus angles of incidence $\alpha_{\rm i}$ and scattering $\alpha_{\rm f}$. In the maps the intensity of the specular reflection is imaged along the diagonal ridge with the coordinates $\alpha_{\rm i}=\alpha_{\rm f}$. Curved side bands traced along the lines where $cos\alpha_{\rm i}-cos\alpha_{\rm f} \approx\pm n\lambda/D_{\rm st}$ with the integer $n$, indicate the evolution of the intensities of Bragg diffraction of the first and the second order. The high intensity close to the origin is due to the contamination by the direct beam at low angles of incidence.
The intensity maps displayed in the right column of Fig.~\ref{fig6} were simulated within the framework of the Distorted Wave Born Approximation~\cite{toperverg02,zabel07} (DWBA), which takes into account the optical effects, e.g. reflection and refraction for both neutron spin components, while the lateral structure is considered as a perturbation. The intensity of the reflectivity ridge is then calculated via convolution of the data by fitting with the instrumental resolution function. Following, the intensities of the Bragg diffraction bands were adjusted by varying the SLD's of neighboring stripes, keeping the mean value of the SLD averaged over the stripes fixed to the value determined for each layer from the reflectivity fit. Firstly, applying this procedure to the saturated samples has shown that in the sample with $D_{\rm st}=4~\mu$m the oxidized stripe fraction~\footnote{For the simulation the exact stripe width of the oxidized stripes was determined to be 1.45~$\mu$m.  The deviation to $D_{\rm st}/2=2~\mu$m results from a systematic offset of about 0.5~$\mu$m due to the lithography process.} has a slightly lowered nuclear SLD and magnetization, reduced by about 10\%. This may be due to some oxidation of the NiFe layers for the unprotected etched stripes. After all structural parameters are fixed, the Bragg intensity simulations are adjusted by varying the tilt angles $\beta_1$ and $\beta_2$ of the neighboring stripe magnetization, while keeping the $\langle\cos\Delta\gamma\rangle$ fixed at the value determined by fitting the specular reflectivity for the respective field values.
The procedure is easily accomplished for the case of homogeneously magnetized stripes, which do not produce diffuse scattering. This is particularly the case for the investigated samples displayed in Fig.~\ref{fig6}, where no appreciable diffuse scattering is found. Thus, the chosen assumptions for the analysis are valid. 

\subsection{Interplay between ferromagnetic film thickness and widths of stripes}

\begin{figure}[htbp]
\includegraphics[width=8cm]{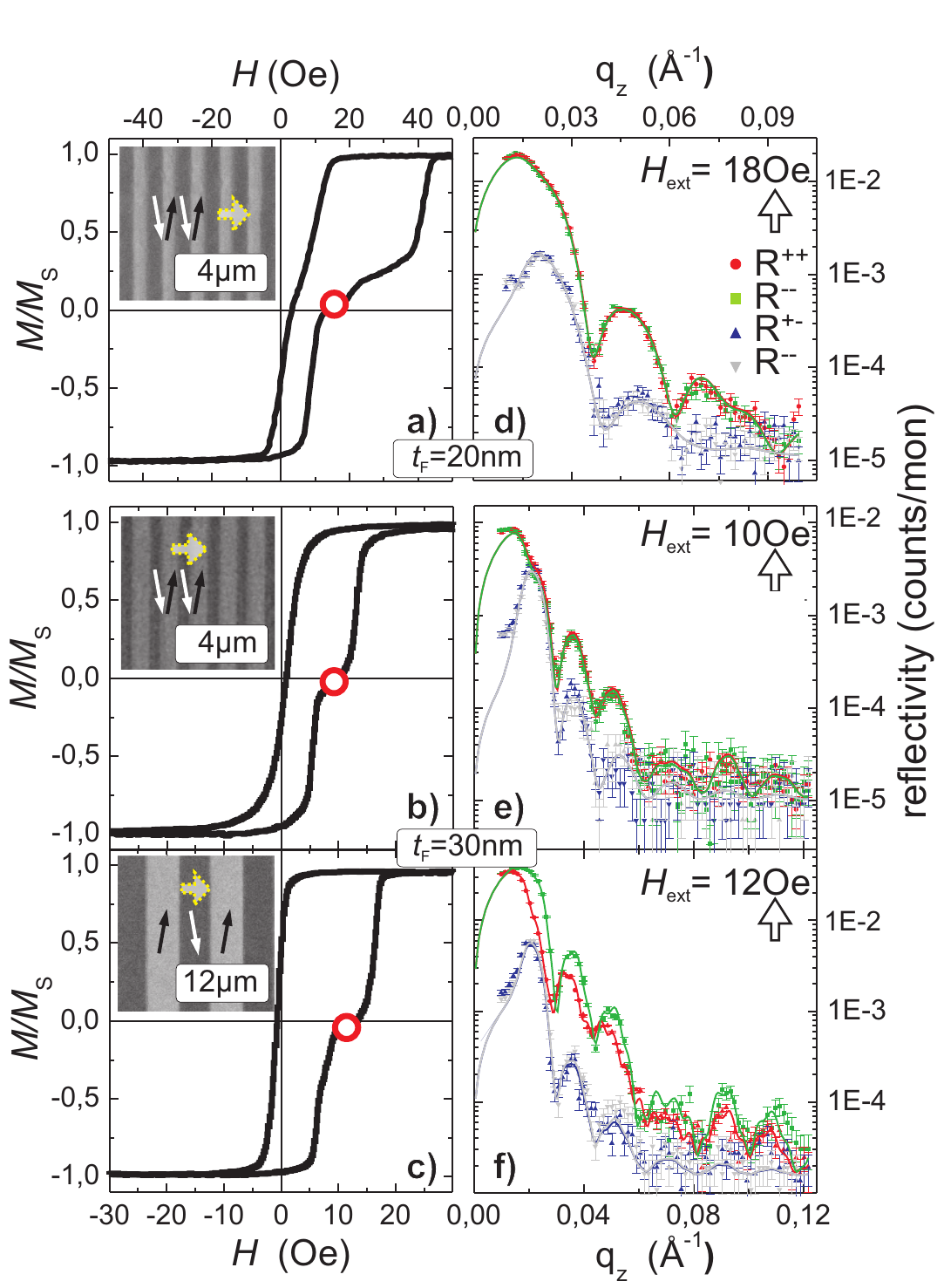}
\caption{\label{fig7} (Color online) Magnetization loops in longitudinal sensitivity in the left column (a-c) with the indicated
magnetic field at the step and representative Kerr image. The according experimental specular reflectivity measurements (symbols) and fits to the data (lines) are displayed in the right column (d-f). In the top row the data for the structures with NiFe of $t_{\rm F}=20~$nm and below the results for $t_{\rm F}=30$~nm are displayed. The magnetic field values and stripe periods of $4~\mu$m and $12~\mu$m are indicated. Note the different scales for the data of the two different ferromagnetic thicknesses $t_{\rm F}$ of 20~nm and 30~nm.}
\end{figure}

To study the influence of strength of exchange bias and direct ferromagnetic exchange coupling across the stripe interfaces, the 4~$\mu$m patterned films with the two different ferromagnetic thicknesses 20~nm and 30~nm were compared with the 30~nm thick film patterned with a period of 12~$\mu$m. In Fig.~\ref{fig7} data collected at an external magnetic field close to the coercive field at the plateau of the loop is shown. The magnetization loops for the three different samples (Fig.~\ref{fig7}, left column) look similar, all exhibiting a magnetization step for fields applied opposite to the exchange bias direction at $M/M_{\rm S}\approx 0$ (see inset in Fig.~\ref{fig7}a-c). Thereby, the thinner F with $t_{\rm F}=20$~nm Fig.~\ref{fig7}a) exhibits the strongest shift with $H_{\rm eb}\approx30$~Oe for the exchange biased stripe fraction. This results in the broader
plateau of about $20$~Oe and leads to the assumption of the largest opening angle of the two stripe magnetizations at the hysteresis step. Comparing the specular reflectivities (Fig.~\ref{fig7}, right column), for the same stripe period $D_{\rm st}=4~\mu$m (Fig.~\ref{fig7}d,e) the frequency of the reflectivity oscillations is altered, comparing the two different F thicknesses. The characteristic shape has not changed. This is due to the F layer thickness, i.e. the mean optical potential, being the largest contribution to the specular intensity. The NSF intensities are equal (SA~$\approx$~0) and the SF is increased. This again indicates stripe magnetizations of opposite direction, but with a strong transverse component of magnetization.\footnote{For $t_{\rm F}=30$~nm at the stripe period of $12~\mu$m (Fig.~\ref{fig6}f) the NSF intensities are uncompensated with $R^{++}< R^{--}$. This is caused by an exchange biased magnetic frame around the sample still magnetized opposite to the magnetic field direction. This frame was present for the neutron measurements but was removed for the inductive magnetization loop measurements. For the fit of the specular intensities, the surrounding frame was included in the
fitting model.}

\begin{figure}[htbp]
\includegraphics[width=6cm]{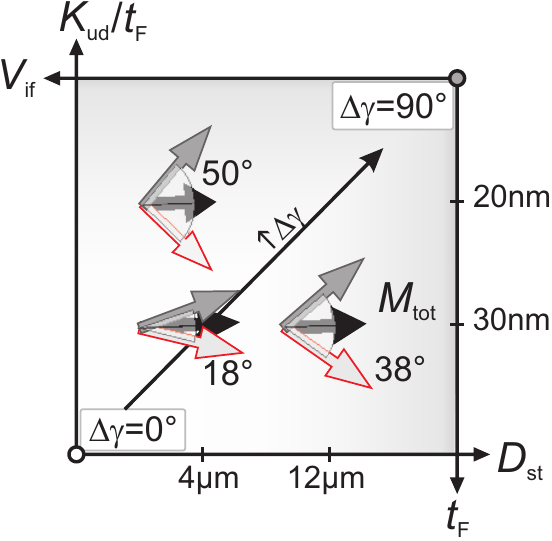} 
\caption{\label{fig8} (Color online) Direction of the net total magnetization $M_{\rm tot}$ and opening angle $\Delta\gamma$ at the partial antiparallel state calculated from the fit to the specular PNR data in \ref{fig7}(d-f). With decreasing $t_{\rm F}$ at constant stripe period the opening angle increases. Likewise, at constant $t_{\rm F}=30~nm$, with increasing stripe period, i.e. reduced F stripe interface volume $V_{\rm if}$, the angle increases. The corresponding stripe magnetization directions are marked by different colors with the EB stripes in red (pointing down) and the free F (NiFe) fraction in dark gray (pointing up) (cf. Fig.~\ref{fig5}).}
\end{figure}

The calculated net magnetization $M_{\rm tot}$, angle $\gamma$, and the magnetization reduction due to the magnetization
splitting of the differently biased stripes are summarized in Figure~\ref{fig8}. For all three samples at the hysteresis step, the net magnetization is canted about $\gamma=87^\circ$ away from the stripe axis. The sample with the lower exchange bias values in the biased stripes with a F thickness of $30$~nm, and the smallest pattern period of $D_{\rm st}=4~\mu$m shows the smallest splitting of neighboring stripe magnetizations of $\Delta\gamma=18^\circ$. By increasing the period to $D_{\rm st}=12~\mu$m and thus reducing the number of stripe interfaces, the perpendicular magnetization component for both stripe types is decreasing. For this case, the splitting angle is $\Delta\gamma=38^\circ$. Keeping the stripe period constant but increasing the effective unidirectional anisotropy ${\rm K_{ud}}/t_{\rm F}$ of the F-AF stripes, $\Delta\gamma$ increases in a similar way, now up to $50^\circ$.

This is in agreement with the phenomenological model proposed in Ref.~\onlinecite{theis-brohl06}, according to which the transversal magnetization component in magnetic thin films with antiparallel exchange bias modulated stripes is proportional to the stripe period, i.e. stripe width, and the F film thickness. Yet, in the model the domain wall width $w_{\rm dw}$ at the stripe interface was assumed to be small with $w_{\rm dw}\ll D_{\rm st}/2$ and was neglected. However, for thin films with $t_{\rm F}\leq 100$~nm the hybrid film domain walls are of the N\'eel type and magneto-static energy terms play a significant role in the development of the magnetization patterns in the intermediate magnetic state of ``oppositely'' aligned magnetization. The N\'eel walls consist of a small wall core and an elongated wall tail to reduce the stray field energy contributions in the thin films. According to Hubert and Sch\"afer\cite{hubert98} the width of the N\'eel wall tails $w_{\rm tail}$ in thin films is approximated by extrapolating the logarithmic wall profile to the point, where the magnetization component along the wall normal is zero ($\cos\theta=$~0). The deduced domain wall tail extending into the neighboring domain is calculated from
\begin{equation}
    w_{\rm tail} \approx  0.56 \cdot t_{\rm F} \cdot \frac{K_{\rm d}}{K_{\rm u,eff}},
\end{equation}
with the stray field anisotropy constant $K_{\rm d}=~J_{\rm S}^2/2~\mu_{\rm 0}$ and the effective uniaxial ferromagnetic anisotropy constant $K_{\rm u,eff}$. Under the assumption that the N\'eel wall positioned in the F (NiFe/IrMnO$_x$) stripe with the lower effective anisotropy and maximal F layer uniaxial anisotropy of $K_{\rm u,F}=~10^3~\rm J/m=~K_{\rm u,eff}$ and the saturation magnetization $J_{\rm S}=~\mu_{\rm 0}M_{\rm S}= 1.04$~T the N\'eel wall tail is determined to be approximately $4.8~\mu$m. Alternatively, considering a domain wall in the exchange biased stripe volume, the effective anisotropy is calculated by $K_{\rm u,eff}=~K_{\rm u,F}+K_{\rm ud,F-AF}/t_{\rm F}$. With $H_{\rm eb}=30$~Oe and thus an additional contribution of $K_{\rm ud,F-AF}/t_{\rm F}=2.5\cdot 10^3~\rm J/m^3$ to the effective anisotropy still a tail of about $1.4 ~\mu$m is expected. 
Comparing the obtained values with the stripe dimensions ($D_{\rm st}/2=~2.6~\mu$m) it is obvious that the N\'eel wall tails extend significantly into the stripe volume. The wall's magnetic tails even overlap and thereby contribute to the occurrence of the residual transverse magnetization and increasing tilt angle for the narrow stripes. Thus, the N\'eel walls themselves play an additional role to the magnetization orientation at the state of oppositely aligned magnetization. Furthermore, the observations of the domain walls with the same magnetization rotation sense (see also Ref.~\onlinecite{theis-brohl06} and Ref.~\onlinecite{mccord05}) fortify the fact that the transverse magnetization signal of the N\'eel walls does not cancel out and consequently contributes to the transverse component. Nevertheless, even with the consideration of a real domain wall structure the general statements of the simplified model, as discussed, above are still valid. 

\subsection{Laterally resolved MOKE magnetometry}

\begin{figure}[htbp]
\includegraphics[width=8cm]{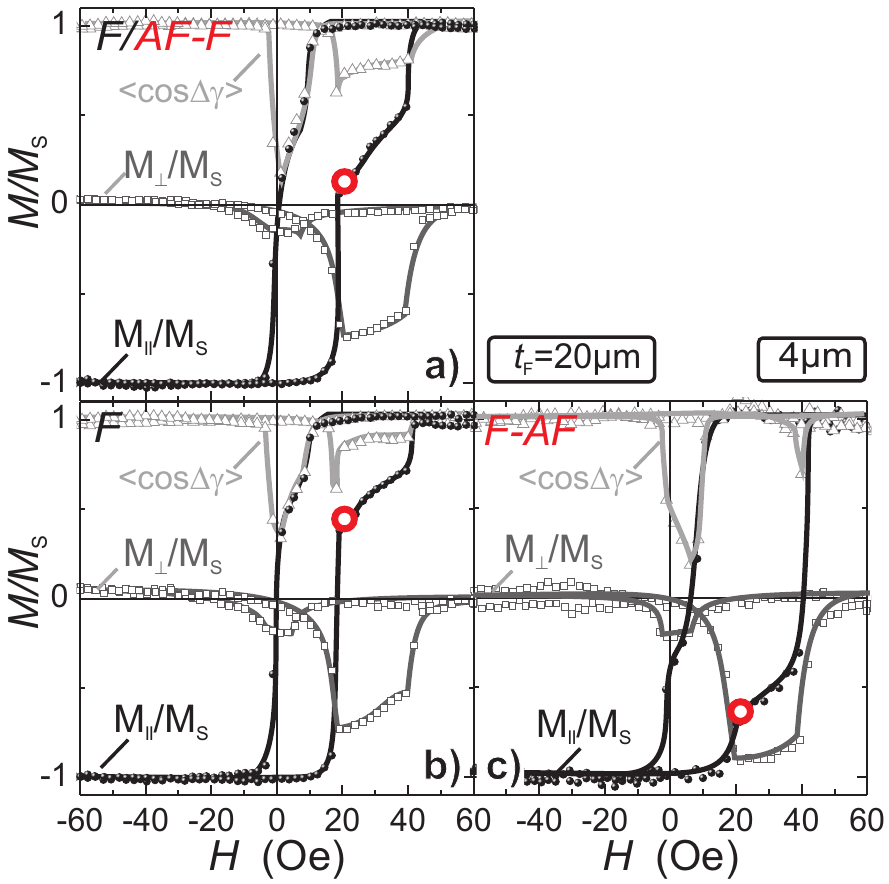}
\caption{\label{fig9} (Color online) Longitudinal ($\parallel$) and transversal ($\perp$) magneto optical magnetization loops of
$\rm NiFe(20~nm)/IrMn-IrMnO/Ta-IrMnO$ the stripe period $D_{\rm st}=~4~\mu$m. The mean Kerr signal averaged over several stripes
(F/AF-F) is shown in a). Local measurements for the free (F) and exchange biased (AF-F) stripes are plotted in b) and c). The symbols represent the experimental data whereas the lines are introduced as guide to the eye. }
\end{figure}

A complementary method to check for magnetization rotation and the occurrence of transversal magnetization components during the stripe reversal is high resolution MOKE magnetometry. By this method, not only the overall magnetization response can be extracted, but the magnetization of each type of stripes can be probed independently. Adjusting the Kerr sensitivity to be either longitudinal ($\parallel$) or transversal ($\perp$) to the stripe, respectively field axis, the mean magnetization along and perpendicular to the stripes can be probed. To have comparable signal amplitudes along the different directions, only the longitudinal Kerr sensitivity contrast was used, while for the measurement of the transversal signal, the sample and magnetic field were rotated by $90^\circ$ simultaneously. The longitudinal, $I_{\parallel}$, and transversal, $I_{\perp}$, intensities were normalized to the saturated stripe intensity $I_{\rm S}$. For the data, the respective longitudinal signal is proportional to the normalized mean magnetization $M_{\parallel}/M_{\rm S,y}=\langle\cos\gamma\rangle$ and thus comparable to the angles deduced from PNR, where $\langle\cos\gamma\rangle$ is a fitting
parameter. Additionally, the reduction of the mean magnetization $\langle\cos\Delta\gamma\rangle$ can be calculated from the Kerr intensities by
\begin{equation}
\langle\cos\Delta\gamma\rangle =\sqrt{\left(\frac{M_{\rm\parallel}}{M_{\rm S}}\right)^2+\left(\frac{M_{\rm \perp}}{M_{\rm S}}\right)^2}\leq 1.
\end{equation}
Exemplary, the magneto optical magnetization loops for the exchange biased patterned stripes (F/AF-F) of the $20~$nm sample are shown in Figure~\ref{fig9}. Averaging over several stripes one obtains a two-step magnetization loop similar to the integral inductive measurement. At the magnetization plateau a strong transverse signal as well as $\langle\cos\Delta\gamma\rangle<1$, hinting for magnetization dispersion, are observed. Due to the different covering layers the optical properties of the two stripe types vary. Therefore, the resulting mean Kerr signal is not only a function of the stripe and wall fractions, but also depends on the locally different magneto-optic sensitivity or contrast. 
By selectively analyzing the magnetic response of the different stripes this limitation can be overcome
(Fig.~\ref{fig9}~b,c). After the partial reversal of the free stripes (Fig.~\ref{fig9}~b) {\itshape both} stripe types do show an increased transversal magnetization signal ($\perp$), which is consistent with the PNR data. Thereby the free stripes' magnetization completely collapses along the magnetic field direction and thus the transverse magnetization diminishes only when the biased stripe fraction is reversed (Fig.~\ref{fig9}~c). The F stripes do not rotate coherently at the plateau, because a reduction of the mean F stripe magnetization by $\langle\cos\Delta\gamma\rangle=0.84<1$, equivalent to $\Delta\gamma= 33^\circ$, is observed. This means, that even after the switching of the F stripe, the magnetization is not homogeneously aligned, but possibly modulated in terms of magnetization ripple which is not resolved in the Kerr images [Fig.~\ref{fig3}d)]. Also, an inhomogeneous magnetization distribution due to formation of the N\'eel wall tails might contribute to the reduction of the magnetic moment. Yet, for the exchange biased stripe (F-AF) the assumption of homogeneous magnetization rotation is confirmed by $\langle\cos\Delta\gamma\rangle=1$. Therefore, the mean magnetization angle $\gamma$ for the F-AF stripes can be calculated from $\cos\gamma\approx 0.69$ to $134^\circ$ (cf. Fig.~\ref{fig9},$\circ$). This is almost equal to the value of
$137^\circ$, which was obtained from the PNR fits with the mean magnetization along $\gamma=87^\circ$ and $\Delta\gamma=50^\circ$. Due to the magnetization dispersion of the F stripes and the largely different averaging volumes of the local magneto optic hysteresis and integral PNR, the other magnetization angles cannot be compared directly. Yet, the local Kerr data affirms the existence of transversal magnetization components, both, in the free F and exchange biased AF-F stripes in the configuration of oppositely aligned magnetization.

\section{Conclusions}

In conclusion, exchange bias patterned films were investigated by complementary inductive and high resolution magneto optic magnetometry, Kerr microscopy observation, and polarized neutron reflectivity measurements to obtain a detailed picture of the magnetization reversal in terms of the competition of interfacial exchange bias and stripe interface contributions. Analyzing the specular neutron reflectivity, the magnetization orientation during the reversal process was extracted. Thereby, also
the existence of an intermixing of the magnetic $\rm NiFe$ with the neighboring layers was revealed. Comparing the magnetization
orientation for different ferromagnetic layer thickness $t_{\rm F}$ and stripe period $D_{\rm st}$ at the magnetization plateau, in the presumed ``antiparallel state'', it is shown that the net magnetization is oriented almost perpendicular to
the stripe axis for all samples. The detailed study confirms the substantial influence of the ferromagnetic layer thickness $t_{\rm F}$ and stripe period $D_{\rm st}$ on the angle between neighboring free F and exchange biased F-AF stripes. Here, the thinner the F layer and higher the effective unidirectional anisotropy contribution $K_{\rm ud}/t_{\rm F}$, the higher is the
angle of magnetization tilting. The tilting increases with increasing stripe period. This emphasizes the significance of the competition of inter-facial exchange bias and inter-stripe ferromagnetic coupling and thus confirms earlier
phenomenological considerations~\cite{theis-brohl06}. However, in contrast to the simple model, an additional contribution of the widely extending N\'eel walls structures to the transverse magnetization orientation is discussed. The transverse
magnetization and thus coupled switching of the stripe magnetizations is complemantary proven by the local magneto optic
magnetometry of the individual stripes. Our investigations unambiguously demonstrate the crucial role of structure dimensions due to interface interactions in hybrid magneto-patterned films. Thus the presented results prove the possibility to tune the magnetization orientation and reversal of exchange bias modulated thin films by means of adjusting the F film thickness as well as structure dimensions. 
\begin{acknowledgments}
We thank R. Sch\"afer and F. Bruessing for valuable discussions and help in the preparation of the manuscript. Founding through the Deutsche Forschungsgemeinschaft and the BMBF (grant no.: BMBF 03ZA7BOC) is highly acknowledged.
\end{acknowledgments}


\end{document}